\documentclass[%
 reprint,
 amsmath,amssymb,
 aps,
 pra,
 showkeys,
floatfix,
]{revtex4-2}
\usepackage[utf8]{inputenc}
\usepackage{graphicx}
\usepackage{dcolumn}
\usepackage{bm}
\usepackage{bbold}
\usepackage{physics}
\usepackage[colorlinks=true, allcolors=blue]{hyperref}
\usepackage{siunitx}
\usepackage{mathdots}
\usepackage[dvipsnames]{xcolor}
\DeclareSIUnit\dBm{dBm}
\begin{document}
\preprint{APS/123-QED}
\title{Nonreciprocal scattering in a microwave frequency comb}%
\author{Christoph L. Bock}%
\author{J. C. Rivera Hernández}%
\author{Fabio Lingua}%
\author{David B. Haviland}\email{haviland@kth.se}%
\affiliation{Department of Applied Physics, KTH Royal Institute of Technology, SE-10691 Stockholm, Sweden}%
\date{\today}

\begin{abstract}
We investigate nonreciprocal scattering within the modes of a microwave frequency comb.
Adjusting the pump frequencies, amplitudes, and phases of a Josephson parametric oscillator, we control constructive interference for the $m \longrightarrow \ell$ scattering processes, while concurrently achieving destructive interference for the inverse process $\ell \longrightarrow m$.
We outline the methodology for realizing nonreciprocity in the context of two-mode isolation and a three-mode circulation, which we extend to multiple modes.
We find good agreement between the experiments and a linearized theoretical model.
Nonreciprocal scattering expands the toolset for parametric control, with the potential to engineer alternative quantum correlations.
\end{abstract}


\maketitle


\newcommand{\matr}[1]{\bm{#1}}
\newcommand{\identity}{\mathbb{1}}
\newcommand*\operator[1]{\hat{#1}}
\newcommand{\Oq}{\operator{Q}}
\newcommand{\Oa}{\operator{A}}
\newcommand{\Ophi}{\operator\Phi}
\newcommand{\rel}{\text{rel}}

\definecolor{drkgreen}{rgb}{0.0, 0.5, 0.0}
\definecolor{violet}{rgb}{0.5, 0.0, 1.0}
\newcommand{\rd}[1]{\textcolor{red}{#1}}
\newcommand{\mgt}[1]{\textcolor{magenta}{#1}}
\newcommand{\gr}[1]{\textcolor{drkgreen}{#1}}
\newcommand{\bl}[1]{\textcolor{blue}{#1}}
\newcommand{\vl}[1]{\textcolor{violet}{#1}}

\section{\label{sec:introduction}Introduction}

Microwave superconducting circuits have emerged as a leading platform for quantum technology~\cite{devoret_superconducting_2013, gu_microwave_2017, arute_quantum_2019, gyenis_moving_2021, bardin_microwaves_2021}.
A measurement of the quantum state of these circuits typically requires amplifiers that introduce excessive backaction and undesired decoherence.
Measurement through nonreciprocal devices mitigates the backaction and is therefore considered essential to any microwave quantum technology.
Achieving nonreciprocity in a compact and reconfigurable measurement scheme, which does not rely on bulky magnetic-based circulators, is highly desired.
Josephson parametric amplifiers (JPAs) can in theory work at the quantum limit of backaction~\cite{caves_quantum_2012}, but their typical mode of operation is reciprocal~\cite{yurke_lownoise_1996, castellanos-beltran_widely_2007, castellanos-beltran_amplification_2008, yamamoto_flux-driven_2008, bergeal_phase-preserving_2010, aumentado_superconducting_2020}.

Significant effort has focused on Josephson-junction devices for directional quantum-limited amplification~\cite{abdo_directional_2013, metelmann_nonreciprocal_2015, macklin_nearquantum-limited_2015, renberg_nilsson_high-gain_2023} and non-magnetic circulation~\cite{kamal_noiseless_2011, estep_magnetic-free_2014, kerckhoff_-chip_2015}.
Nonreciprocal scattering was demonstrated in a passive Josephson-junction ring \cite{fedorov_nonreciprocity_2024}, as well as nonreciprocal frequency conversion and amplification in a multimodal Josephson circuit~\cite{lecocq_nonreciprocal_2017, lecocq_microwave_2020, lecocq_efficient_2021}.
These solutions often require multiple coupled resonators and other components which constrain efficient signal routing, bandwidth, and tunability.
Simplifying the circuitry while maintaining nonreciprocal behavior for tunable, selected bandwidths would greatly enhance the scalability and integration of quantum hardware.

This work explores the use of a single JPA to nonreciprocally connect input and output tones in a microwave frequency comb.
We analyze nonreciprocity in the context of asymmetric scattering between these tones.
We can calculate the scattering matrix elements using coupled mode theory in terms of directional digraphs~\cite{deak_reciprocity_2012, ranzani_graph-based_2015}.
Multiple digraphs connecting two modes create the possibility of interference, described by a loop phase analogous to the phase acquired by charged particles circulating a loop enclosing magnetic flux~\cite{yuan_synthetic_2018, fang_realizing_2012}.
In our experiments, we control this loop phase through the amplitude and phase of multiple low- and high-frequency pumps, each locked to the common phase reference of the tones in the microwave comb.
We explore different nonreciprocal scattering regimes, ranging from two-mode isolation and three-mode circulation to a general nonreciprocal scattering between $n=41$ modes.
Our scattering approach represents a first step toward a compact, scalable, and reconfigurable nonreciprocal measurement scheme for microwave quantum technology.

\section{Setup and Experiment} \label{sec:experiment}

\begin{figure}[h]
    \centering
    \includegraphics[width=0.95\linewidth]{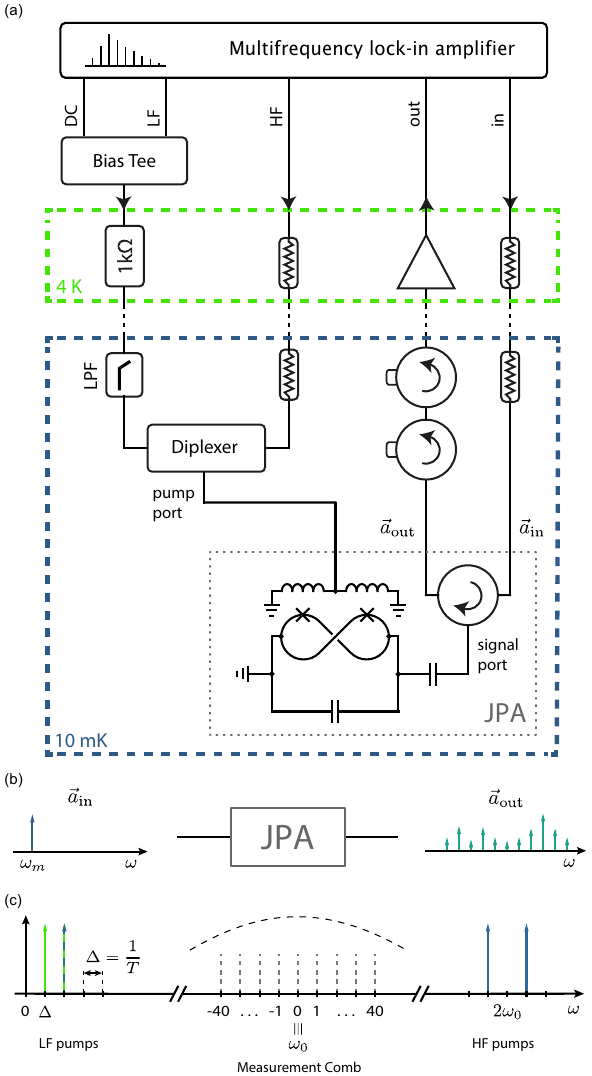}
    \caption{(a) Simplified experimental setup, neglecting multiple filtering stages at intermediate temperatures in the dilution refrigerator. 
    The JPA's resonant frequency is modulated by an external flux gradient threading a gradiometric SQUID loop.
    (b) Protocol for the scattering experiment: for each input tone at frequency $\omega_m$, the response is measured simultaneously across all modes in the defined frequency comb, capturing the full scattering profile of the JPA.
    (c) Frequency-domain representation of the measurement basis and the pump configurations. Blue arrows denote the isolator pumping scheme with $\Omega_{1,2}=2\omega_0\pm \Delta$ and $\Omega_3=2\Delta$; green arrows denote the circulator pumping scheme with $\Omega_1=\Delta$ and $\Omega_2=2\Delta$.
    }
    \label{fig:jpa-schem}
\end{figure}

The experiments measure the scattering of a weak input signal that intermodulates with a multifrequency pump waveform, producing multiple intermodulation products called idlers.
The JPA consists of a lumped-element $LC$ circuit, incorporating a gradiometric superconducting quantum interference device (SQUID) whose inductance is modulated by an external time-dependent signal applied through the pump port [see Fig.~\ref{fig:jpa-schem}(a)].
A dc bias at the pump port tunes the resonant frequency of the JPA to $\omega_0 = 2\pi \times \SI{4.2}{\giga\hertz}$.
The signal port is overcoupled, yielding a loaded quality factor $Q=37.5$, corresponding to a linewidth $\kappa = 2 \pi \times \SI{112}{\mega\hertz}$.
A circulator separates the input and output signals, directing the output to a cryogenic low-noise amplifier.

A digital multifrequency lock-in amplifier~\cite{tholen_measurement_2022} generates the pump tones and the weak coherent signal while demodulating at up to 192 frequencies.
We chose all pump, signal, and demodulation tones from a mode basis $\{a_m\}$ whose frequencies are integer multiples of the measurement bandwidth $\Delta = 1/T = 2 \pi \times \SI{125}{\kilo\hertz}$, where $T$ is the measurement time window.
This choice ensures orthogonality of the modes $\{a_m\}$ and establishes a global phase reference for modulation and demodulation.
We further eliminate Fourier leakage between demodulated modes by tuning $\Delta$ to be commensurate with the digital sampling frequency.
The enumeration scheme for the mode frequencies is $\omega_m = \omega_0 \pm m\Delta$, $m \in \mathbb{N}$, with the center frequency $\omega_0$ placed within $\Delta$ of the resonant frequency of the JPA [see Fig.~\ref{fig:jpa-schem}(c) for details].

We synthesize the multifrequency pump waveform by superposing tones selected from the same basis $\{a_m\}$, distinguishing between low- and high-frequency pump tones.
Low-frequency pumps have $\Omega_k = k \Delta$, while high-frequency pumps have $\Omega_k = 2\omega_0 \pm k \Delta$, where $k$ is a small positive integer [see Fig.~\ref{fig:jpa-schem}(c) for details].
The synchronous low- and high-frequency pumps are output through different ports of the multifrequency lock-in amplifier and combined at the mixing chamber using a diplexer, as shown in Fig.~\ref{fig:jpa-schem}(a).

In Fig.~\ref{fig:jpa-schem}(b) we sketch the protocol used to measure the scattering matrix.
We send a single coherent tone at frequency $\omega_m$ of approximately $140$ photons, properly thermalized at \SI{10}{mK}, through the input port of the JPA, while measuring the scattered frequency comb in the output.
We step the single coherent tone, repeating the process for every mode of the measurement basis.

\section{Nonreciprocal scattering} \label{sec:theory}

For small perturbations of the magnetic-flux gradient, the effective Hamiltonian describes the JPA dynamics
\begin{equation}\label{eq:HJPA}
    \hat{\mathcal{H}}_\text{JPA} = \frac{\Ophi^2}{2L_0} + \frac{\Oq^2}{2C} + \frac{\Ophi^2}{2\var{L(t)}},
\end{equation}
where operators $\Oq$, and $\Ophi$ represent the node charge and flux at the JPA's input port, respectively.
We express the Hamiltonian in~\eqref{eq:HJPA} in terms of the canonical ladder operators of the unperturbed harmonic oscillator~\cite{yamamoto_principles_2016}, 
\begin{equation}\label{eq:HJPA-2}
        \hat{\mathcal{H}}_\text{JPA} = \frac{\omega_0}{2}\left(A^\dag A + AA^\dag\right) + \frac{\omega_0}{2}p_L(t)\left(A + A^\dag\right)^2,
\end{equation}
which results in the standard single harmonic-oscillator Hamiltonian, plus a time-dependent quadratic term modulated by the pump signal $p_L(t)$.
The pump waveform is defined as a sum of coherent tones with amplitude $p_k$, frequency $\Omega_k$, and phase $\phi_k$,
\begin{equation}\label{eq:pump}
    p_L(t)=\sum_k p_k\cos{(\Omega_kt + \phi_k)}.
\end{equation}

We expand Eq.~\eqref{eq:HJPA-2} in a basis of discrete modes with frequency $\omega_m$ corresponding to traveling waves in the transmission line which are orthogonal in the measurement time interval $T$: $A^{(\dag)}=\sum_m a_m^{(\dag)}e^{-i\omega_m t}$.
When the pump signal takes the form~\eqref{eq:pump}, the time-dependent problem described by Eq.~\eqref{eq:HJPA-2} can be mapped to the problem of $n$-coupled harmonic oscillators (see Ref.~\cite{rivera_hernandez_control_2024} for the full derivation and discussion).
Considering the classical limit $a^{(\dag)}_m\rightarrow a_m^{(*)}\in\mathbb{C}$, one recovers a linear system of equations of motion 
\begin{equation} \label{eq:EOM}
    \dv{a_m}{t} = -i\Tilde{\omega}_0 a_m -i\omega_0 p_L(t)(a_m + a^*_m) + \sqrt{\gamma}\, a_{\text{in,}m}
\end{equation}
where $\Tilde{\omega}_0 = \omega_0 - i\frac{\gamma}{2}$ and $\gamma$ accounts for losses due to coupling to the transmission line (here assumed constant for all modes).
In the frequency domain the system~\eqref{eq:EOM} is usually expressed in matrix form $-i\gamma\matr{M}\vec{a} = \sqrt{\gamma} \vec{a}_{\text{in}}$, where $\vec{a}=(a_1, a_1^*, \dots, a_m, a_m^*, \dots)$ is the vector containing the complex mode amplitudes. We normalize $\matr{M}$ by $\gamma$ as in Ref.~\cite{ranzani_graph-based_2015}.
The solution of Eq.~\eqref{eq:EOM_f} is then readily available upon inversion of matrix $\matr{M}$~\cite{naaman_synthesis_2022}.

The scattering matrix follows from the mode-matching condition $\sqrt{\gamma}\vec{a} = \vec{a}_{\text{in}} + \vec{a}_{\text{out}}$ between the internal degrees of freedom, and the external incoming and reflected signals 
\begin{equation} \label{eq:S}
   \vec{a}_\text{out} = \matr{S} \cdot \vec{a}_\text{in} = \left( i\matr{M}^{-1} - \identity \right) \vec{a}_\text{in},
\end{equation}
where $\gamma$ is the coupling to the transmission line, assumed to be constant for all modes (see Ref.~\cite{rivera_hernandez_control_2024} for a detailed derivation).
Each element $S_{mn}\neq 0$ connects the input mode at frequency $\omega_n$ to the output mode at $\omega_m$ through pump-controlled mixing processes.

The pumps enable second-, third-, and higher-order intermodulation (wave-mixing) products, which are the idlers in the output signal.
When multiple pumps are involved, different mixing processes interfere at the same idler frequency, effectively enhancing or canceling scattering to that frequency, dependent on the amplitudes and phases of the pumps.
Carefully engineering the pump frequencies, amplitudes, and phases, we exploit the interference properties of overlapping mixing processes to create nonreciprocal scattering.
Nonreciprocity emerges as an asymmetry of the scattering matrix $\matr{S} \neq \matr{S}^T$~\cite{ranzani_graph-based_2015}, i.e. $S_{ij} \neq S_{ji}$ holds for at least a subset of matrix elements.
Here we use a stronger definition of nonreciprocity $|\matr{S}| \neq |\matr{S}^T|$, thus excluding nonreciprocal phase shifts.

We know of no universal recipe for generating a specific nonreciprocity, but a necessary condition is that the pump frequencies satisfy $\Omega_p = \sum_{k \neq p} n_k \Omega_k$, where $n_k \in \mathbb{N}$ is the multiplicity of pump $k$, accounting for higher-order mixing processes~\cite{lecocq_nonreciprocal_2017}.

\subsection{Two-mode nonreciprocal scattering: the isolator}

\begin{figure*}[htbp]
    \centering
    \includegraphics[width=\textwidth]{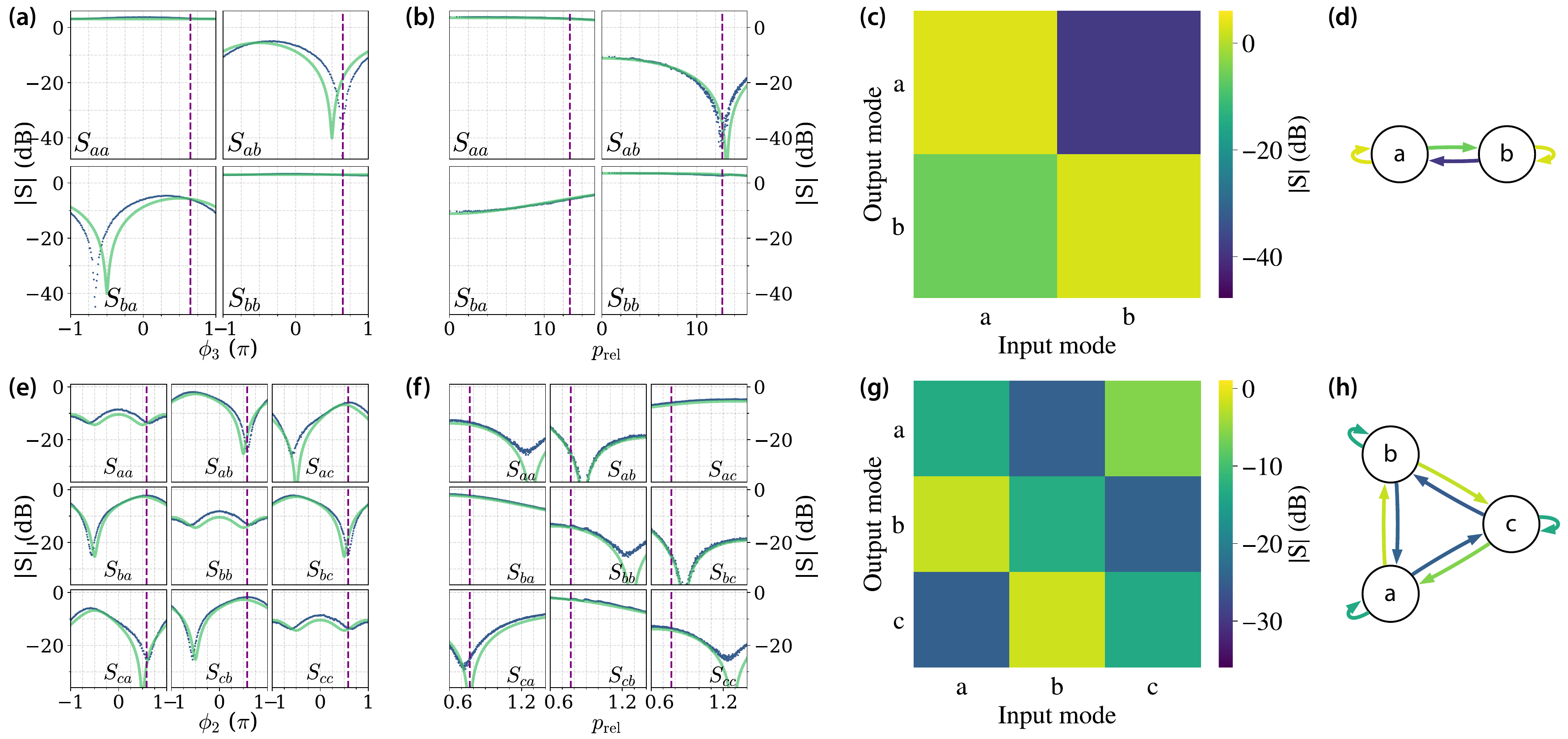}
    \caption{(a)-(d) Isolator. (e)-(h) Circulator. (a)~and~(e) Scattering parameters as a function of the phase $\phi_3$ and $\phi_2$ for fixed pump powers, respectively. (b)~and~(f) Scattering parameters as a function of the relative amplitude $p_\rel$ for fixed phases. Blue dots refer to experimental data and green lines to numerical solutions. (c)~and~(g) Measured scattering matrix at the optimal amplitude and phase values given by the vertical dashed lines. (d)~and~(h) Graph representation of the scattering matrix.}
    \label{fig:main}
\end{figure*}

Isolation is achieved with two high-frequency pumps at ${\Omega_1 = 2\omega_0 - \Delta}$ and ${\Omega_2 = 2\omega_0 + \Delta}$, together with a low-frequency pump at ${\Omega_3 = 2 \Delta}$ [see blue arrows Fig.~\ref{fig:jpa-schem}(c)].
These pump detunings, which are arbitrary provided that ${\Omega_2 = \Omega_3 + \Omega_1}$, achieve isolation between mode pairs $\omega_a$ and $\omega_b = \omega_a + 2 \Delta$.
Figure~\ref{fig:main}(a) shows the $S$ parameters between such a pair of modes, measured as a function of the phase of the low-frequency pump $\phi_3$.
Figure~\ref{fig:main}(b) shows the dependence of the $S$ parameters on the amplitude of the low-frequency pump relative to that of the high-frequency pumps $p_\rel = p_{3} / p_{1,2}$.
The high-frequency pumps have phase zero and the same amplitude, which is chosen to get \SI{3}{\dB} signal gain.

The diagonal elements $\abs{S_{ii}}$ are weakly dependent on $\phi_3$ and $p_\rel$.
At the optimal phase $\phi_3 = \SI{0.65}{\pi}$ and pump amplitude $p_\rel = 12.7$ simultaneous destructive and constructive interference gives about \SI{39}{\dB} of isolation and \SI{6}{\dB} of insertion loss.
Reprogramming the pump waveform to $\phi_3=\SI{-0.67}{\pi}$ reverses the direction of isolation, without physically changing the circuit or connection ports.
In Fig.~\ref{fig:main}(c) we plot the scattering matrix at the optimal values $\phi_3$ and $p_\rel$.
The directed graph in Fig.~\ref{fig:main}(d) summarizes the isolator, with arrows color-coded according to the $S$ parameters.

We find qualitative agreement between the experimental results and the numerical solution of the equations of motion, as shown by the green lines in Fig.~\ref{fig:main}(a)~and~(b).
We find a discrepancy between the experimental and theoretical value of the phase $\phi_3$ for destructive interference, that we attribute to nonlinearity in the experiment which is not accounted for in the theoretical model.
We analyze the interference process between $S_{ab}$ and $S_{ba}$ using the graphical Cramer method~\cite{ranzani_circulators_2019, ranzani_graph-based_2015} (see Appendix~\ref{appA} for details).
For the isolator, the mode couplings, determined by the pumping scheme, are illustrated in Fig.~\ref{fig:digraph-iso}(a), with a detailed discussion provided in Appendix~\ref{appA_iso}.

The exact expansion of $S_{ab}$ and $S_{ba}$ is shown in Fig.~\ref{fig:digraph-iso}(b).
The diagonal elements of the matrix $\matr{M}$, denoted as $\Delta_m = \frac{1}{\gamma}\left(-\omega_m - \omega_0 + i\gamma/2\right) = \kappa_m e^{i\alpha_m}$, encode information about the frequency and losses of the mode $m$, while $-\Delta_m^*$ encode the anti-mode.
$S_{ab}=0$ arises from the destructive interference between the second-order intermodulation product involving the pump at $\Omega_3$ and the third-order intermodulation products involving the pumps at $\Omega_1$ and $\Omega_2$.

To match the experimental conditions, we assume weak pumps with equal strengths for the high-frequency pumps ($|g_1| = |g_2| = g\ll1$) and a relative strength for the low-frequency pump ($|g_3| = r g$, with $r \in \mathbb{R}$).
Setting $S_{ab} = 0$ we arrive at the condition for perfect destructive interference,
\begin{align}
    g^2=rg\kappa_d \equiv \rho \label{eq:iso-cond1}, \\
    \phi_{loop} = -\alpha_d\label{eq:iso-cond2}
\end{align}
where $\phi_{loop} = \phi_1-\phi_2 + \phi_3$ is the total phase of the loop between the three interfering pumps shown in blue in Fig.~\ref{fig:digraph-iso}(a).
In our experiment we tune $\phi_1=\phi_2$ resulting in $\phi_{loop}=\phi_3= -\alpha_d$.
Substituting the conditions ~\eqref{eq:iso-cond1}~and~\eqref{eq:iso-cond2} into the expression for $S_{ba}$, we find $S_{ba} \propto \rho \left(1 - e^{-2i\alpha_d}\right)$, which reaches a maximum when $\phi_3=\frac{\pi}{2}$.
The analytic expansion thus explains the nonreciprocal behavior observed in the experiment, which shows good agreement with the numerical solutions of the equations of motion.
\begin{figure}
    \centering
    \includegraphics[width=\columnwidth]{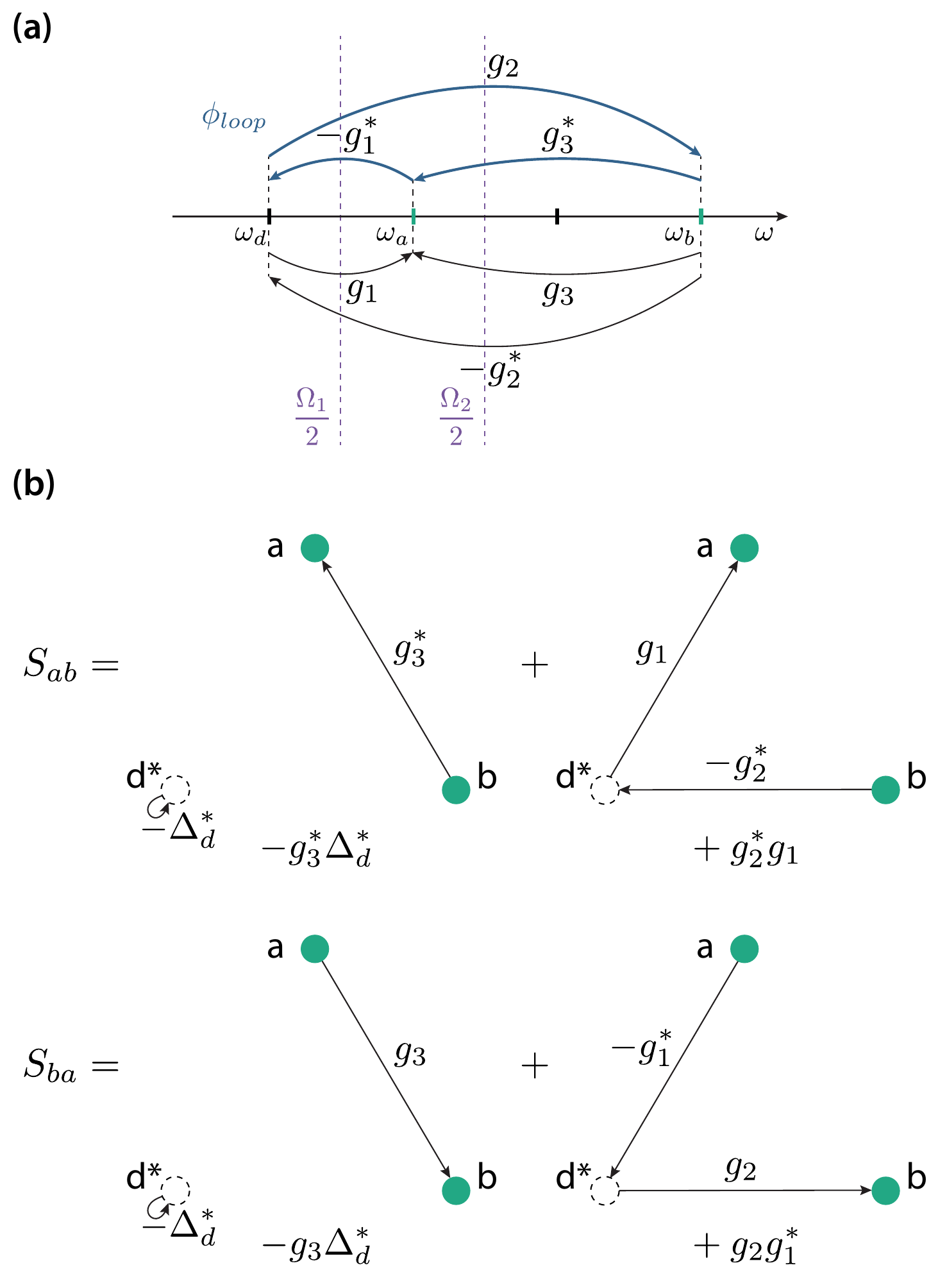}
    \caption{Isolator mode coupling scheme (a), digraph expansion of scattering matrix elements $S_{ab}$, $S_{ba}$ (b).}
    \label{fig:digraph-iso}
\end{figure}

\subsection{Three-mode nonreciprocal scattering: the circulator}

Circulation is realized with two low-frequency pumps at $\Omega_1 = \Delta$ and $\Omega_2 = 2 \Delta$, which couple three modes at frequencies $\omega_b$, $\omega_a = \omega_b - \Delta$, and $\omega_c = \omega_b + \Delta$ [see green arrows Fig.~\ref{fig:jpa-schem}(c)].
The interference between different orders of intermodulation generating idlers at these three frequencies depends on the phase of either pump and their relative amplitude.
Figure~\ref{fig:main}(e) shows the scattering as a function of $\phi_2$ at a fixed value of $p_\text{rel} = p_1 / p_2 = 0.77$.
At $\phi_2 = \SI{0.58}{\pi}$, constructive interference enhances $S_{ba}$, $S_{ac}$, and $S_{cb}$, while destructive interference suppresses $S_{ca}$, $S_{ab}$ and $S_{bc}$, establishing clockwise circulation between the modes, $a \rightarrow b \rightarrow c \rightarrow a$.

In Fig.~\ref{fig:main}(f) we sweep the relative pump amplitude while fixing $\phi_2 = \SI{0.58}{\pi}$.
At $p_\rel = 0.77$, we achieve balanced isolation $S_{ca},\, S_{ab},\, S_{bc} \simeq \SI{-24}{\dB}$, insertion loss $S_{ac},\, S_{ba},\, S_{cb} \simeq \SI{-3}{\dB}$, and reflection $S_{aa},\, S_{bb},\, S_{cc} \simeq \SI{-13}{\dB}$.
At $p_\rel = 1.24$ we minimize reflection to \SI{-25}{\dB} at the cost of reducing and unbalancing isolation to $S_{ca} \simeq \SI{-10}{\dB}$ and $S_{ab},\, S_{bc} \simeq \SI{-19}{\dB}$.
Figure~\ref{fig:main}(g) illustrates the scattering matrix for balanced isolation.
Figure~\ref{fig:main}(h) summarizes the circulation in a directed graph.
Changing $\phi_2$ to $\SI{-0.58}{\pi}$ reverses the direction of circulation.

\begin{figure}
    \centering
    \includegraphics[width=\columnwidth]{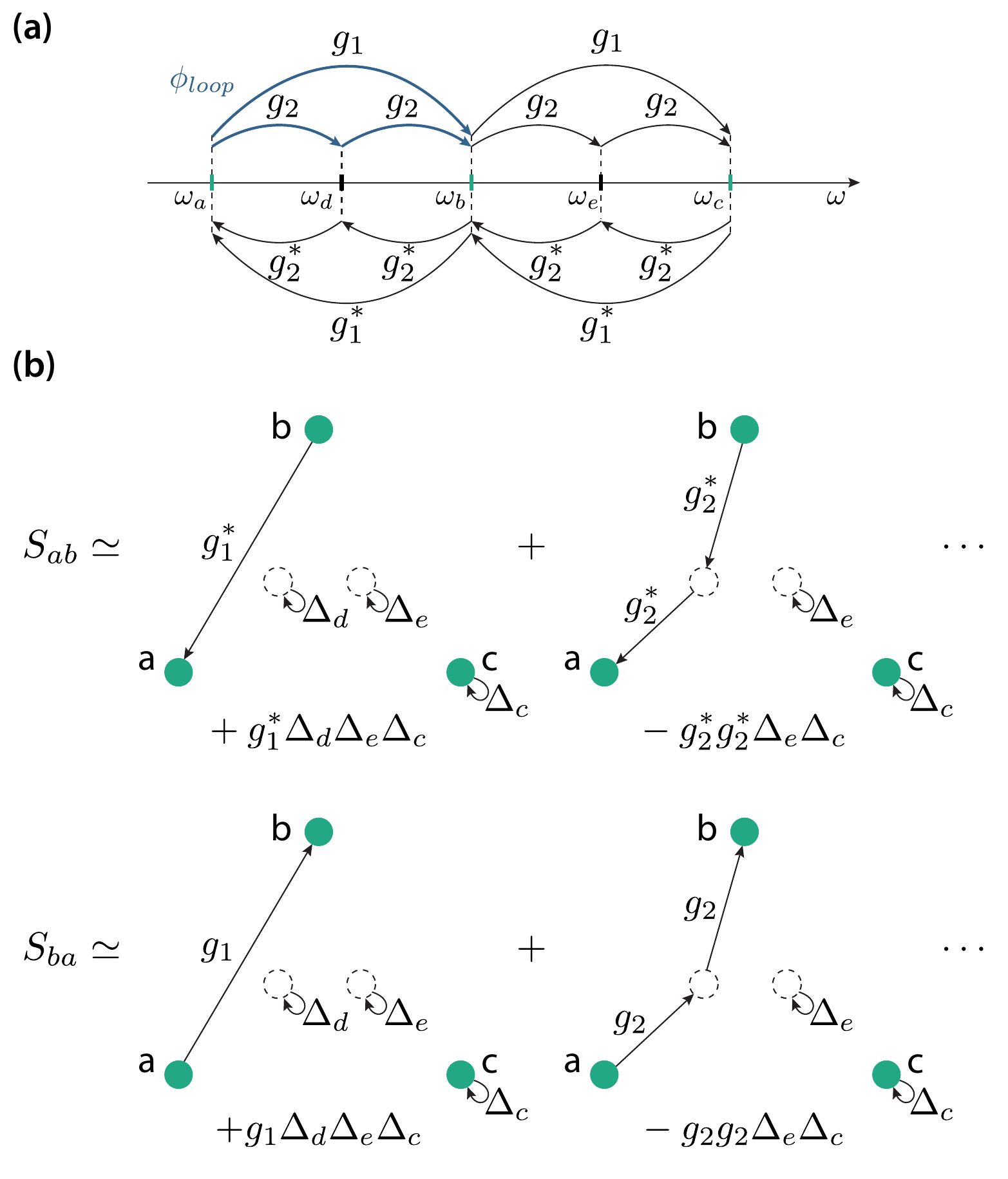}
    \caption{Circulator mode-coupling scheme (a), second-order digraph expansion of scattering matrix elements $S_{ab}$, $S_{ba}$ (b).}
    \label{fig:digraph-circ}
\end{figure}

The theoretical model shows similarly good agreement with the experimental results [see green lines in Fig.~\ref{fig:main}(e)~and~(f)].
To understand the mixing processes underlying the circulator's nonreciprocal behavior, we consider the five modes depicted in Fig.~\ref{fig:digraph-circ}(a).
With $n=5$ modes the full analytical Cramer method becomes impractical.
However, approximate expressions for $S_{ab}$ and $S_{ba}$ can be derived by truncating the digraph expansion to higher-order terms in pump power. Figure~\ref{fig:digraph-circ}(a) shows the couplings provided by the two low-frequency pumps. 
We define the couplings as $g_1 = g e^{i/\phi_1}$ and $g_2 = rg e^{i/\phi_2}$, with $r = \frac{|g_1|}{|g_2|} \in \mathbb{R}$, and assume $g \ll 1$ to match the weak pump regime of our experiment (see appendix~\ref{appA_circ}).
Figure~\eqref{fig:digraph-circ}(b) shows the digraph expansion of $S_{ab}$ and $S_{ba}$ to second order in $g$, corresponding to the expressions
\begin{multline}
    S_{ab}\simeq \frac{i}{|M|}(\; ge^{-i\phi_1}\Delta_d\Delta_e\Delta_c +\\
    - r^2g^2e^{-i2\phi_2}\Delta_e\Delta_c + o(g^3)\;),
    \label{eq:circ-Sab}
\end{multline}
\begin{multline}
    S_{ba}\simeq \frac{i}{|M|}\big(\; ge^{i\phi_1}\Delta_d\Delta_e\Delta_c +\\
    - r^2g^2e^{i2\phi_2}\Delta_e\Delta_c + o(g^3) \;\big).
    \label{eq:circ-Sba}
\end{multline}
This expansion shows that the relevant interfering process comes from mixing products that involve the pump $\Omega_1$ and twice the pump $\Omega_2$.
Taking $S_{ab}=0$ we retrieve the conditions on the magnitude $g\kappa_d=r^2g^2 \equiv \rho$ and on the total phase of the loop of $\phi_{loop} =2\phi_2-\phi_1= - \alpha_d$ [blue loop in Fig.~\ref{fig:digraph-circ}(a)].
Setting $\phi_1=0$ we derive 
\begin{equation}
    S_{ba}\propto \rho\left(e^{i\alpha_d} - e^{-i\alpha_d}\right)=2i\rho\sin\alpha_d\neq0,
\end{equation}
which is generally nonzero because the diagonal elements $\Delta_m$ contain the losses $\gamma$ due to coupling to the transmission line.
Similar arguments can be extended to higher-order terms ($O(g^3)$) to prove nonreciprocal scattering between modes $a$ and $c$.
Details are provided in Appendix~\ref{appA_circ}.

This analytical approximation confirms the circulation behavior shown in the experimental and numerical results. 
Whenever the condition $S_{ab}=S_{bc}\simeq S_{ca}\simeq 0$ holds, $S_{ba}=S_{cb}\simeq S_{ac}\neq 0$ holds.

\subsection{$n$-mode nonreciprocal scattering}

We extend the isolator and circulator schemes to achieve nonreciprocal scattering across $n=41$ modes.
We measure the scattering of a coherent tone stepped through the mode basis $\{ a_m \}$ while simultaneously listening at all frequencies.

\begin{figure}
    \centering
    \includegraphics[width=\linewidth]{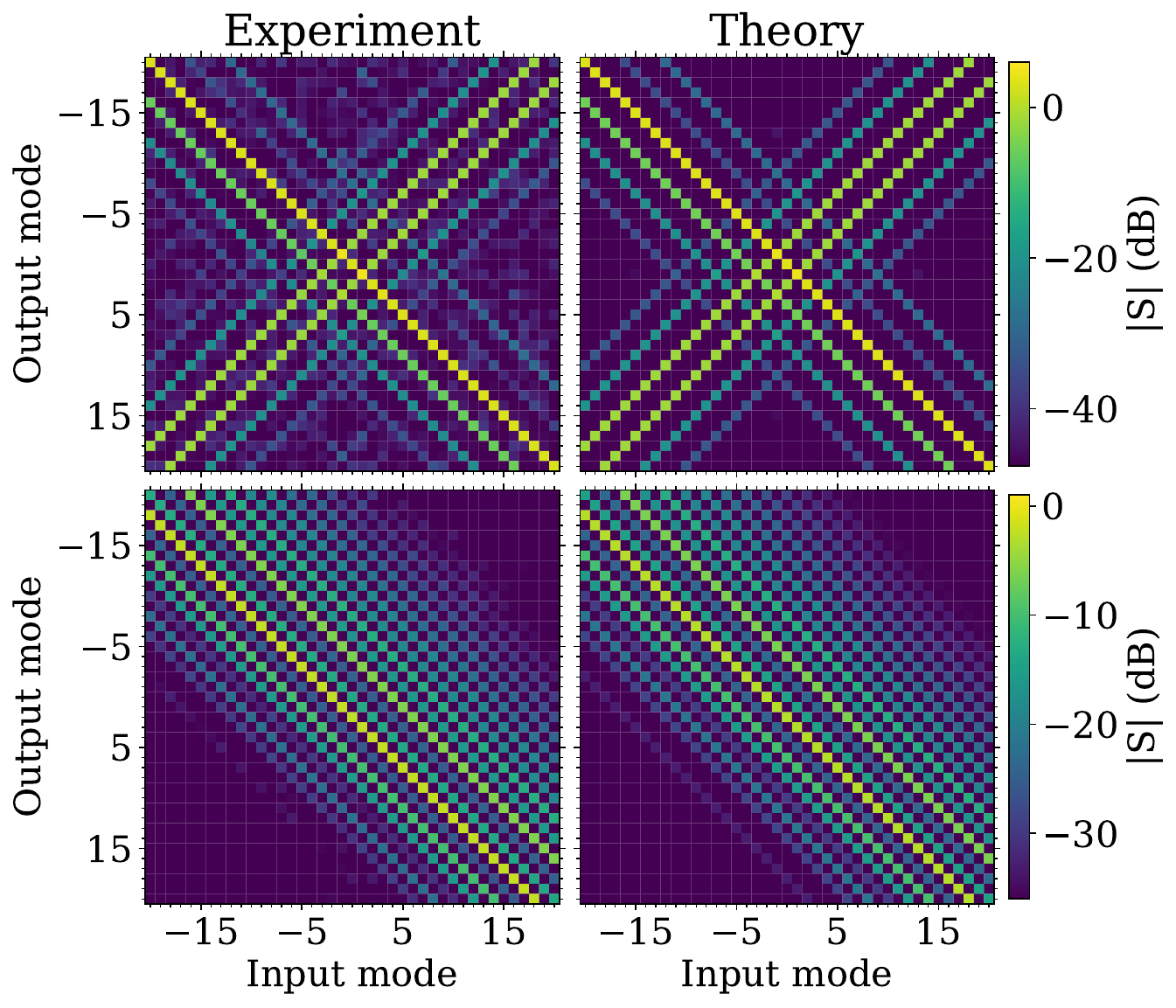}
    \caption{Experimental (left) and theoretical (right) scattering matrices. 
    The upper panels correspond to pumping the JPA with two high-frequency tones at $\Omega_{1,2}=2\omega_0\pm2\Delta$ and one low-frequency pump at $\Omega_3=4\Delta$.
    The lower panels correspond to pumping the JPA with two low-frequency pumps at $\Omega_1=2\Delta$ and $\Omega_2=4\Delta$.
    The magnitude of the scattering is expressed in dB, normalized to the pump-off case.}
    \label{fig:Smat}
\end{figure}

Top panels of Fig.~\ref{fig:Smat} show the measured scattering matrix, normalized to the zero-pump case, for two high-frequency pumps at ${\Omega_{1,2} = 2\omega_0 \pm 2\Delta}$, and a low-frequency pump at ${\Omega_3 = 4 \Delta}$, as in the two-mode isolator configuration.
We fix the pump amplitudes and phases according to the optimal parameters in the two-mode isolator [see vertical dashed lines in Fig.~\ref{fig:main}(a)~and~(b)] to cancel the first upper diagonal in $\matr{S}$, allowing transmission along the lower diagonal.
Adjusting the phase to $\phi_3 = \SI{-0.67}{\pi}$ cancels the lower diagonal while transmitting the upper diagonal.

Bottom panels of Fig.~\ref{fig:Smat} show the normalized $\matr{S}$ matrix for two low-frequency pumps at $\Omega_1 = 2 \Delta$ and $\Omega_2 = 4 \Delta$, following the three-mode circulator scheme.
We set the pump amplitudes and phases according to the optimal parameters in the three-mode circulator [see vertical dashed lines in Fig.~\ref{fig:main}(e)~and~(f)].
The $\matr{S}$ matrix reveals an imbalance between the upper and lower diagonals, demonstrating nonreciprocity.
However, we do not observe circulation between all $n=41$ modes, likely due to increased complexity and overlapping scattering pathways.

Theoretical scattering matrices align well with the experimental results, as shown in Fig.~\ref{fig:Smat}.
We find optimal parameters for the effective transmission line coupling $\gamma$, pump amplitudes $p_k$, and pump phases $\phi_k$ that minimize the sum of the element-wise square distance between the numerical and measured scattering matrices.

\section{\label{sec:conclusion}Conclusion}

We studied the nonreciprocal scattering of a JPA in different regimes. 
We defined nonreciprocity in terms of asymmetry in scattering magnitude between input and output frequency modes. 
Pumping the JPA with a combination of low- and high-frequency pumps, we realized different nonreciprocal behavior ranging from two-mode isolation to three-mode circulation.
The presence of multiple pumps creates a multitude of three- and four-wave mixing processes (second- and third-order intermodulation) connecting the input signal to a given idler.
By carefully tuning pump frequencies, phases, and amplitudes we create destructive interference for processes $m\longrightarrow \ell$ while simultaneously creating constructive interference for the inverse $\ell \longrightarrow m$ processes.
We provided a general recipe for nonreciprocity extending the case of a two-mode isolator to a measurement basis of $n=41$ modes, with good agreement between experiment and theory.
This approach constitutes another building block in the toolset of parametric control, with the ability to easily reconfigure nonreciprocal mode scattering.

\begin{acknowledgments}
We acknowledge Joe Aumentado and John Teufel at the National Institute of Standards and Technology (NIST) for helpful discussions and for providing the JPA used in this experiment.  We also acknowledge fruitful discussions with Kirill Petrovnin, Ilari Lilja, Ekaterina Mukhanova, and Pertti Hakonnen at Aalto University.
This work was supported by the Knut and Alice Wallenberg Foundation through the Wallenberg Center for Quantum Technology (WACQT).
\end{acknowledgments}

\section*{Author Declaration}

C. L. B. performed both the experiment and the numerical analysis. J. C. R. H. conceived the main idea and performed the initial investigation. F. L. performed the theoretical analysis. D. B. H. supervised the work. All the authors discussed, edited, and reviewed the manuscript.

D. B. H. is part owner of the company Intermodulation Products AB, which produces the digital microwave platform used in this experiment.

\section*{Data Availability}
The data that support the findings of this study are openly available in Zenodo at~\cite{bock_data_2025}.

\appendix

\section{Analytical analysis of scattering} \label{appA}

The equations of motion \eqref{eq:EOM} in the frequency domain take the form
\begin{multline}
    \label{eq:EOM_f}
   i\left(\omega_m + \Tilde{\omega}_0\right) a_m  +\\ + i\sum_n\left(g_{mn}a_m + g_{mn}^*a^*_m\right) = \sqrt{\gamma}\, a_{\text{in,}m}
\end{multline} 
where the couplings $g_{mn}=\sum_k\Tilde{g}_k \delta_{m,n\pm k}$ between modes $m$ and $n$ depend on the sum of the pump complex amplitude $\Tilde{g}_k=\omega_0 p_ke^{i\phi_k}$.
Here $\delta_{m,n\pm k}$ selects modes whose frequencies satisfy $\omega_m=\omega_n \pm \Omega_k$, where $\Omega_k$ is the pump frequency. 
We refer to Ref.~\cite{rivera_hernandez_control_2024} for a full derivation of this theoretical framework.
As mentioned in the main text, the equations of motion \eqref{eq:EOM} can be written in the matrix form
\begin{equation} \label{EOM_matrix}
    -i\gamma\matr{M} \vec{a} = \sqrt{\gamma} \vec{a}_\text{in},
\end{equation}
where we normalize $\matr{M}$ by $\gamma$ as in Ref.~\cite{ranzani_graph-based_2015}.
The matrix $\matr{M}$ is a $2n \times 2n$ matrix and takes the general form
\begin{equation}
    \matr{M} = \matr{M}_0 + \sum_k \matr{G}_k.
\end{equation}
Here $\matr{M}_0$ is a $2n \times 2n$ diagonal matrix containing the mode frequencies 
\begin{equation}
    \matr{M}_0 = \begin{pmatrix}
            \Delta_{-\frac{n}{2}} & & & & \\[7 pt]
            & -\Delta_{-\frac{n}{2}}^* & & & \\
            & & \ddots & & \\
            & & & \Delta_\frac{n}{2} & \\[7 pt]
            & & & & -\Delta_\frac{n}{2}^* \\
        \end{pmatrix},
\end{equation}
where an element $\Delta_m = \frac{1}{\gamma}\left(-\omega_{m} -\tilde{\omega}_0\right) \in \mathbb{C}$, while the $\matr{G}_k$ are $2n \times 2n$ coupling matrices due to pump $k$.
The nonzero elements $m\ell$ of each matrix $\matr{G}_k$ couple modes $m$ and $\ell$ through pump $k$, in direct proportion to the complex amplitude $g_k = \frac{|\Tilde{g}_k|}{\gamma} e^{i\phi_k}$. Note that while the $\Tilde{g}_k$ have units of frequency, the rescaled $g_k$ in the matrix representation of the equations of motion are nondimensional.
For low-frequency pumps, where $\Omega_k = k\Delta$, we have
\begin{equation}
    \matr{G}_k = \begin{pmatrix}
        0 & & & g_k & & & & & \\[5 pt]
        & 0 & & & -g_k^* & & & & \\
         & & \ddots & & & \ddots & & & \\
         g_k^* & & & 0 & & & g_k & & \\[5 pt]
         & -g_k & & & 0 & & & -g_k^* \\
         & & \ddots & & & \ddots & & \\
         & & & g_k^* & & & 0 & \\[5 pt]
         & & & & -g_k & & & 0 \\
    \end{pmatrix},
\end{equation}
with two nonzero off-diagonals at position $\pm k$.
For high-frequency pumps, where $\Omega_k = 2\omega_0 + k\Delta$, we have 
\begin{equation}
    \matr{G}_k = \begin{pmatrix}
             0 & & & & & & g_k & & \\[3 pt]
             & 0 & & & & -g_k^* & & & \\[-2 pt]
             & & \ddots & & \iddots & & & & \\
             & & \iddots & & \ddots & & & & \\
             & g_k & & & & 0 & & & \\[3 pt]
             -g_k^* & & & & & & 0 & & \\[-4 pt]
             & & & & & & & \ddots & \\[-2 pt]
             & & & & & & & & 0 \\
        \end{pmatrix},
\end{equation}
with one nonzero antidiagonal in position $k$.

Solutions to Eq.~\eqref{EOM_matrix} are readily available by inverting the matrix $\matr{M}$, with the scattering matrix given by Eq.~\eqref{eq:S}.
For a small number of modes, one can analytically compute $\matr{M}^{-1}$ following the graphical Cramer method used in Refs.~\cite{ranzani_graph-based_2015, ranzani_circulators_2019}.
This diagrammatic method computes the determinant of $\matr{M}$ and the elements of $\matr{M}^{-1}$ by summing weighted directed graphs, called \emph{digraphs}.
Digraph vertices represent modes, while edges correspond to mode couplings induced by the pumps.

The determinant of $\matr{M}$ is computed by summing \emph{all} possible digraphs, while the matrix element $M^{-1}_{m\ell}$ is the sum of those digraphs with a path from mode $\ell$ to $m$ (see Fig.~\ref{fig:digraph-iso} for an example).
The elements of the scattering matrix are
\begin{multline}  \label{eq:Smk}
    S_{m\ell} = \frac{i}{|\matr{M}|} \sum_{ \ell \rightarrow m}^{\text{paths}} (-1)^{2n+l} M_{\ell, i_1} \dots M_{i_{n-4}, m} - \delta_{m \ell},
\end{multline}
where the sum runs over all the possible paths from mode $\ell$ to $m$, each with $2n-1$ edges.
The paths are defined by the product of edges $M_{\ell, i_1} M_{i_2,i_3} \dots M_{i_{n-4}, m}$ through permutation of the indices $\{ m, i_1, \dots i_{2n-4}, \ell \}$, $i_j \in \{ [1,2n], i_j \neq m, \ell \}$.
The sign is determined by the exponent $l$ referring to the number of loops (closed paths) in the digraph.

Exact computation beyond a few modes is infeasible due to factorial growth of the number of digraphs $N_g=(2n)!$.
Even the minimal cases considered here are challenging.
However, truncating the sum to the most relevant digraphs offers insights into the dominant mixing processes responsible for nonreciprocity.

\subsection{The isolator} \label{appA_iso}

To explain the nonreciprocal scattering of the isolator one has to consider at least three modes: the two modes $\omega_a = \omega_0$ and $\omega_b = \omega_0 + 2\Delta$, and an auxiliary mode $\omega_d = \omega_0 - \Delta$.
The two high-frequency pumps couple modes symmetrically around half their pump frequency by coupling their respective anti-modes, while the low-frequency pump couples modes separated by the pump's frequency, as shown in Fig.~\ref{fig:digraph-iso}(a).

The dynamics of the three modes is described by the $6 \times 6$ matrix $\matr{M}$.
An analytical approach using Eq.~\eqref{eq:Smk} would involve $N_g=6!=720$ digraphs.
We distinguish two independent subsets that are not coupled to each other: the set of equations of modes $a_a$, $a_b$, and anti-mode $ a_d^*$ and the set of equations of anti-modes $ a_a^*$, $ a_b^*$, and mode $a_d$.
By focusing on one subset, the dimensionality of the problem reduces, simplifying $\matr{M}$ to
\begin{equation}
   \matr{M} = \begin{pmatrix}
            -\Delta_d^* & g_1 & g_2 \\
             -g_1^* & \Delta_a & g_3 \\
             -g_2^* & g_3^* & \Delta_b \\     
        \end{pmatrix}.
        \label{eq:Miso}
\end{equation}

In Fig.~\ref{fig:digraph-iso}(b) we use Eq.~\eqref{eq:Miso} to graphically compute the digraphs contributing to the scattering terms $S_{ab}$ and $S_{ba}$.
For consistency with our experimental setup, we assume equal pump strengths $|g_1|=|g_2|=g$ for the high-frequency pumps and a relative value $|g_3|=rg$ for the low-frequency pump, with $r \in \mathbb{R}$.
The mode couplings become $g_1=ge^{i\phi_1}$, $g_2=ge^{i\phi_2}$ and $g_3=rge^{i\phi_3}$, resulting in
\begin{align}
    S_{ab} &= \frac{i}{|M|} \left( g^2 e^{-i\phi_2} e^{i\phi_1} -rg e^{-i\phi_3} \Delta_d^* \right),
    \label{eq:Sab-iso} \\
    S_{ba} &= \frac{i}{|M|} \left( g^2 e^{i\phi_2} e^{-i\phi_1} - rg e^{i\phi_3} \Delta_d^* \right).
    \label{eq:Sba-iso}
\end{align}
Imposing $S_{ab}=0$ and assuming $\Delta_d^*=\kappa_d e^{-i\alpha_d}$, we derive the equation
\begin{equation}
     g^2 e^{-i\phi_2} e^{i\phi_1} = r g e^{-i\phi_3}\kappa_d e^{-i\alpha_d}.
     \label{eq:Sab_0}
\end{equation}
We then retrieve the conditions
\begin{align}
    &g^2 = rg\kappa_d \equiv \rho \label{eq:cond1}, \\
    &\phi_{loop} = -\alpha_d\label{eq:cond2}
\end{align}
where $\phi_{loop } =\phi_1 - \phi_2 + 
\phi_3$ is the total phase of the loop among the three pumps, as shown in blue in Fig.~\ref{fig:digraph-iso}(a).
In our experiment we set $\phi_1=\phi_2$, leading to $\phi_3=-\alpha_d$.
Plugging the conditions~\eqref{eq:cond1}~and~\eqref{eq:cond2} into Eq.~\eqref{eq:Sba-iso} we find
\begin{equation}
    S_{ba} \propto \rho \left( 1 - e^{-2i\alpha_d} \right) \neq 0,
\end{equation}
which is generally nonzero, as $\alpha_d \neq 0$ due to the presence of losses and coupling in our transmission-line setup ($\gamma\neq0$), proving nonreciprocity since $S_{ab}=0$, $S_{ba}\neq0$.

Finally, adding mode $\omega_e = \omega_0 + \Delta$ does not alter the validity of the results but would increase the dimensionality of $\matr{M}$ and the number of digraphs.
However, it is always possible to retrieve an approximate expression for $S_{ab}$ and $S_{ba}$ identical to Eqs.~\eqref{eq:Sab-iso}~and~\eqref{eq:Sba-iso} by truncating the digraph expansion to the second order in pump power $g$, adopting the same approach shown for the circulator case below. 

\subsection{The circulator} \label{appA_circ}

The minimum number of modes to explain the nonreciprocal scattering of the circulator discussed in this paper is $n=5$ [see Fig.~\ref{fig:digraph-circ}(a)].
With five modes, the total number of digraphs is $N_g = (2 \times 5)! = 3628800$.
However, in this case, the pumping scheme involves only low-frequency pumps $\Omega_1=\Delta$ and $\Omega_2=2\Delta$ rendering the set of equations of motion of modes $a_m$ and anti-modes $a_m^*$ completely decoupled to each other.
Without loss of precision, one can therefore half the number of equations.
The matrix $\matr{M}$ restricted only to the modes becomes 
\begin{equation}
   \matr{M} = \begin{pmatrix}
            \Delta_a & g_2 & g_1 & & \\
             g_2^* & \Delta_d & g_2 & g_1 & \\
             g_1^* & g_2^* & \Delta_b & g_2 & g_1 \\
             & g_1^* & g_2^* & \Delta_e & g_2 \\
             & & g_1^* & g_2^* & \Delta_c \\       
        \end{pmatrix}
    \label{eq:Mcirc}
\end{equation}
The total number of digraphs is now significantly reduced to $N_g=5!=120$, but the full analytical method is still largely impractical. 
Let us redefine the coupling of the pumps as $g_1 = g e^{i\phi_1}$ and $g_2 = rg e^{i\phi_2}$ with $r = \frac{|g_1|}{|g_2|} \in \mathbb{R}$.
Within the weak pump regime used in this work, we assume both $|g_1|,|g_2| \ll 1$ (i.e. $|\Tilde{g}_{1,2}|/\gamma \ll 1$), which implies $g \ll 1$.
We use the analytical Cramer method to estimate the approximate expressions of the matrix elements $S_{ab}$ and $S_{ba}$ by truncating the sum over the digraphs at second order in pump power $g$.
\begin{multline}
    S_{ab}\simeq \frac{i}{|M|}(\; ge^{-i\phi_1}\Delta_d\Delta_e\Delta_c +\\
    - r^2g^2e^{-i2\phi_2}\Delta_e\Delta_c + o(g^3)\;)
    \label{eq:Sab_circ}
\end{multline}
\begin{multline}
    S_{ba}\simeq \frac{i}{|M|}\big(\; ge^{i\phi_1}\Delta_d\Delta_e\Delta_c +\\
    - r^2g^2e^{i2\phi_2}\Delta_e\Delta_c + o(g^3) \;\big)
    \label{eq:Sba_circ}
\end{multline}
Figure~\ref{fig:digraph-circ}(b) shows the relevant digraphs participating in the expression of $S_{ab}$ and $S_{ba}$.
Asking for $S_{ab}=0$ leads to the condition
\begin{equation}
    ge^{-i\phi_1}\kappa_d e^{i\alpha_d} = r^2g^2e^{-i2\phi_2}
    \label{eq:Sab_circ_0}
\end{equation}
where we assume $\Delta_d=\kappa_d e^{i\alpha_d}$ as for the isolator case.
Equation~\eqref{eq:Sab_circ_0} leads to the conditions 
\begin{equation}
    g\kappa_d=r^2g^2 \equiv \rho,\;\;\;\;\;\; \phi_{loop} = - \alpha_d\label{eq:condcirc}
\end{equation}
where $\phi_{loop}=2\phi_2-\phi_1$ is the total phase of the loop between the three interfering pumps shown in blue in Fig.~\ref{fig:digraph-circ}(a).
Inserting~\eqref{eq:condcirc} into~\eqref{eq:Sba_circ} and setting $\phi_1=0$ to match our experiment conditions we get
\begin{equation}
    S_{ba}\propto \rho\left(e^{i\alpha_d} - e^{-i\alpha_d}\right)=2i\rho\sin\alpha_d\neq0, 
    \label{eq:Sba_neq0}
\end{equation}
which is generally different from zero as $\alpha_d \neq 0$ for finite coupling $\gamma$.
We have thus proved the nonreciprocal scattering between modes $a$ and $b$.
Nonreciprocity between modes $c$ and $b$ follows by symmetry of the coupling scheme.

The nonreciprocity between the elements $S_{ac}$ and $S_{ca}$ is less obvious.
As direct coupling between modes $a$ and $c$ is missing in the considered pumping scheme, the explanation requires extending the digraph expansion to terms $O(g^3)$.
\begin{multline}
    S_{ac}\simeq \frac{i}{|M|}\big( -g^2e^{-2i\phi_1}\Delta_d\Delta_e\; + \\
    +r^2g^3e^{-i\phi_1}e^{-2i\phi_2}(\Delta_d + \Delta_b +\Delta_e) + o(g^4) \;\big)
    \label{eq:Sac}
\end{multline}
\begin{multline}
    S_{ca}\simeq \frac{i}{|M|}\big(\; -g^2e^{2i\phi_1}\Delta_d\Delta_e\; + \\
    +r^2g^3e^{i\phi_1}e^{2i\phi_2}(\Delta_d + \Delta_b +\Delta_e) + o(g^4) \;\big)  
    \label{eq:Sca}
\end{multline}
Forcing $S_{ca}=0$ leads to the condition
\begin{equation}
    g^2e^{2i\phi_1}\kappa_{de}e^{i\alpha_{de}} = r^2g^3e^{i\phi_1}e^{2i\phi_2}\kappa_{T}e^{i\alpha_{T}}
\end{equation}
where we define $\Delta_d\Delta_e=\kappa_{de}e^{i\alpha_{de}}$ and $\Delta_d + \Delta_b +\Delta_e=\kappa_{T}e^{i\alpha_{T}}$.
Reshuffling the equation we obtain 
\begin{equation}
    g^2\kappa_{de}e^{i(2\phi_2 - \phi_1)} = r^2g^3\kappa_{T}e^{i(\alpha_{T} - \alpha_{de})}
\end{equation}
which leads to the conditions
\begin{equation}
    g^2\kappa_{de}=r^2g^3\kappa_{T}\equiv\rho^\prime
    \label{eq:Sca_cond1}
\end{equation}
\begin{equation}
    \phi_{loop}=-(\alpha_{T} - \alpha_{de})
    \label{eq:Sca_cond2}
\end{equation}
It is straightforward to see that when condition~\eqref{eq:Sca_cond2} is met $S_{ac}\neq0$.
Furthermore, if $\alpha_{T} - \alpha_{de}\approx\alpha_d$ condition~\eqref{eq:Sca_cond2} coincides with condition~\eqref{eq:condcirc} that sets $S_{ab} \simeq 0$.
This confirms $S_{ab}=S_{bc}\simeq S_{ca}\simeq 0$ yielding full circulation.

\bibliographystyle{unsrt}
\bibliography{main.bib}

\begin{thebibliography}{10}

\bibitem{devoret_superconducting_2013}
M.~H. Devoret and R.~J. Schoelkopf.
\newblock Superconducting {Circuits} for {Quantum} {Information}: {An} {Outlook}.
\newblock {\em Science}, 339(6124):1169--1174, March 2013.

\bibitem{gu_microwave_2017}
Xiu Gu, Anton~Frisk Kockum, Adam Miranowicz, Yu-xi Liu, and Franco Nori.
\newblock Microwave photonics with superconducting quantum circuits.
\newblock {\em Physics Reports}, 718-719:1--102, November 2017.

\bibitem{arute_quantum_2019}
Frank Arute, Kunal Arya, Ryan Babbush, Dave Bacon, Joseph~C. Bardin, Rami Barends, Rupak Biswas, Sergio Boixo, Fernando G. S.~L. Brandao, David~A. Buell, Brian Burkett, Yu~Chen, Zijun Chen, Ben Chiaro, Roberto Collins, William Courtney, Andrew Dunsworth, Edward Farhi, Brooks Foxen, Austin Fowler, Craig Gidney, Marissa Giustina, Rob Graff, Keith Guerin, Steve Habegger, Matthew~P. Harrigan, Michael~J. Hartmann, Alan Ho, Markus Hoffmann, Trent Huang, Travis~S. Humble, Sergei~V. Isakov, Evan Jeffrey, Zhang Jiang, Dvir Kafri, Kostyantyn Kechedzhi, Julian Kelly, Paul~V. Klimov, Sergey Knysh, Alexander Korotkov, Fedor Kostritsa, David Landhuis, Mike Lindmark, Erik Lucero, Dmitry Lyakh, Salvatore Mandrà, Jarrod~R. McClean, Matthew McEwen, Anthony Megrant, Xiao Mi, Kristel Michielsen, Masoud Mohseni, Josh Mutus, Ofer Naaman, Matthew Neeley, Charles Neill, Murphy~Yuezhen Niu, Eric Ostby, Andre Petukhov, John~C. Platt, Chris Quintana, Eleanor~G. Rieffel, Pedram Roushan, Nicholas~C. Rubin, Daniel Sank, Kevin~J.
  Satzinger, Vadim Smelyanskiy, Kevin~J. Sung, Matthew~D. Trevithick, Amit Vainsencher, Benjamin Villalonga, Theodore White, Z.~Jamie Yao, Ping Yeh, Adam Zalcman, Hartmut Neven, and John~M. Martinis.
\newblock Quantum supremacy using a programmable superconducting processor.
\newblock {\em Nature}, 574(7779):505--510, October 2019.
\newblock Publisher: Nature Publishing Group.

\bibitem{gyenis_moving_2021}
András Gyenis, Agustin Di~Paolo, Jens Koch, Alexandre Blais, Andrew~A. Houck, and David~I. Schuster.
\newblock Moving beyond the {Transmon}: {Noise}-{Protected} {Superconducting} {Quantum} {Circuits}.
\newblock {\em PRX Quantum}, 2(3):030101, September 2021.
\newblock Publisher: American Physical Society.

\bibitem{bardin_microwaves_2021}
Joseph~C. Bardin, Daniel~H. Slichter, and David~J. Reilly.
\newblock Microwaves in {Quantum} {Computing}.
\newblock {\em IEEE Journal of Microwaves}, 1(1):403--427, January 2021.

\bibitem{caves_quantum_2012}
Carlton~M. Caves, Joshua Combes, Zhang Jiang, and Shashank Pandey.
\newblock Quantum limits on phase-preserving linear amplifiers.
\newblock {\em Physical Review A}, 86(6):063802, December 2012.
\newblock Publisher: American Physical Society.

\bibitem{yurke_lownoise_1996}
B.~Yurke, M.~L. Roukes, R.~Movshovich, and A.~N. Pargellis.
\newblock A low‐noise series‐array {Josephson} junction parametric amplifier.
\newblock {\em Applied Physics Letters}, 69(20):3078--3080, November 1996.

\bibitem{castellanos-beltran_widely_2007}
M.~A. Castellanos-Beltran and K.~W. Lehnert.
\newblock Widely tunable parametric amplifier based on a superconducting quantum interference device array resonator.
\newblock {\em Applied Physics Letters}, 91(8):083509, August 2007.

\bibitem{castellanos-beltran_amplification_2008}
M.~A. Castellanos-Beltran, K.~D. Irwin, G.~C. Hilton, L.~R. Vale, and K.~W. Lehnert.
\newblock Amplification and squeezing of quantum noise with a tunable {Josephson} metamaterial.
\newblock {\em Nature Physics}, 4(12):929--931, December 2008.
\newblock Publisher: Nature Publishing Group.

\bibitem{yamamoto_flux-driven_2008}
T.~Yamamoto, K.~Inomata, M.~Watanabe, K.~Matsuba, T.~Miyazaki, W.~D. Oliver, Y.~Nakamura, and J.~S. Tsai.
\newblock Flux-driven {Josephson} parametric amplifier.
\newblock {\em Applied Physics Letters}, 93(4):042510, July 2008.

\bibitem{bergeal_phase-preserving_2010}
N.~Bergeal, F.~Schackert, M.~Metcalfe, R.~Vijay, V.~E. Manucharyan, L.~Frunzio, D.~E. Prober, R.~J. Schoelkopf, S.~M. Girvin, and M.~H. Devoret.
\newblock Phase-preserving amplification near the quantum limit with a {Josephson} ring modulator.
\newblock {\em Nature}, 465(7294):64--68, May 2010.
\newblock Publisher: Nature Publishing Group.

\bibitem{aumentado_superconducting_2020}
Jose Aumentado.
\newblock Superconducting {Parametric} {Amplifiers}: {The} {State} of the {Art} in {Josephson} {Parametric} {Amplifiers}.
\newblock {\em IEEE Microwave Magazine}, 21(8):45--59, August 2020.

\bibitem{abdo_directional_2013}
Baleegh Abdo, Katrina Sliwa, Luigi Frunzio, and Michel Devoret.
\newblock Directional {Amplification} with a {Josephson} {Circuit}.
\newblock {\em Physical Review X}, 3(3):031001, July 2013.
\newblock Publisher: American Physical Society.

\bibitem{metelmann_nonreciprocal_2015}
A.~Metelmann and A.~A. Clerk.
\newblock Nonreciprocal {Photon} {Transmission} and {Amplification} via {Reservoir} {Engineering}.
\newblock {\em Physical Review X}, 5(2):021025, June 2015.
\newblock Publisher: American Physical Society.

\bibitem{macklin_nearquantum-limited_2015}
C.~Macklin, K.~O’Brien, D.~Hover, M.~E. Schwartz, V.~Bolkhovsky, X.~Zhang, W.~D. Oliver, and I.~Siddiqi.
\newblock A near–quantum-limited {Josephson} traveling-wave parametric amplifier.
\newblock {\em Science}, 350(6258):307--310, October 2015.
\newblock Publisher: American Association for the Advancement of Science.

\bibitem{renberg_nilsson_high-gain_2023}
Hampus Renberg~Nilsson, Anita Fadavi~Roudsari, Daryoush Shiri, Per Delsing, and Vitaly Shumeiko.
\newblock High-{Gain} {Traveling}-{Wave} {Parametric} {Amplifier} {Based} on {Three}-{Wave} {Mixing}.
\newblock {\em Physical Review Applied}, 19(4):044056, April 2023.
\newblock Publisher: American Physical Society.

\bibitem{kamal_noiseless_2011}
Archana Kamal, John Clarke, and M.~H. Devoret.
\newblock Noiseless non-reciprocity in a parametric active device.
\newblock {\em Nature Physics}, 7(4):311--315, April 2011.
\newblock Publisher: Nature Publishing Group.

\bibitem{estep_magnetic-free_2014}
Nicholas~A. Estep, Dimitrios~L. Sounas, Jason Soric, and Andrea Alù.
\newblock Magnetic-free non-reciprocity and isolation based on parametrically modulated coupled-resonator loops.
\newblock {\em Nature Physics}, 10(12):923--927, December 2014.
\newblock Publisher: Nature Publishing Group.

\bibitem{kerckhoff_-chip_2015}
Joseph Kerckhoff, Kevin Lalumière, Benjamin~J. Chapman, Alexandre Blais, and K.~W. Lehnert.
\newblock On-{Chip} {Superconducting} {Microwave} {Circulator} from {Synthetic} {Rotation}.
\newblock {\em Physical Review Applied}, 4(3):034002, September 2015.

\bibitem{fedorov_nonreciprocity_2024}
Arkady Fedorov, N.~Pradeep Kumar, Dat~Thanh Le, Rohit Navarathna, Prasanna Pakkiam, and Thomas~M. Stace.
\newblock Nonreciprocity and {Circulation} in a {Passive} {Josephson}-{Junction} {Ring}.
\newblock {\em Physical Review Letters}, 132(9):097001, February 2024.
\newblock Publisher: American Physical Society.

\bibitem{lecocq_nonreciprocal_2017}
F.~Lecocq, L.~Ranzani, G.~A. Peterson, K.~Cicak, R.~W. Simmonds, J.~D. Teufel, and J.~Aumentado.
\newblock Nonreciprocal {Microwave} {Signal} {Processing} with a {Field}-{Programmable} {Josephson} {Amplifier}.
\newblock {\em Physical Review Applied}, 7(2):024028, February 2017.

\bibitem{lecocq_microwave_2020}
F.~Lecocq, L.~Ranzani, G.A. Peterson, K.~Cicak, A.~Metelmann, S.~Kotler, R.W. Simmonds, J.D. Teufel, and J.~Aumentado.
\newblock Microwave {Measurement} beyond the {Quantum} {Limit} with a {Nonreciprocal} {Amplifier}.
\newblock {\em Physical Review Applied}, 13(4):044005, April 2020.
\newblock Publisher: American Physical Society.

\bibitem{lecocq_efficient_2021}
F.~Lecocq, L.~Ranzani, G.~A. Peterson, K.~Cicak, X.~Y. Jin, R.~W. Simmonds, J.~D. Teufel, and J.~Aumentado.
\newblock Efficient {Qubit} {Measurement} with a {Nonreciprocal} {Microwave} {Amplifier}.
\newblock {\em Physical Review Letters}, 126(2):020502, January 2021.
\newblock Publisher: American Physical Society.

\bibitem{deak_reciprocity_2012}
L.~Deák and T.~Fülöp.
\newblock Reciprocity in quantum, electromagnetic and other wave scattering.
\newblock {\em Annals of Physics}, 327(4):1050--1077, April 2012.

\bibitem{ranzani_graph-based_2015}
Leonardo Ranzani and José Aumentado.
\newblock Graph-based analysis of nonreciprocity in coupled-mode systems.
\newblock {\em New Journal of Physics}, 17(2):023024, February 2015.
\newblock Publisher: IOP Publishing.

\bibitem{yuan_synthetic_2018}
Luqi Yuan, Qian Lin, Meng Xiao, and Shanhui Fan.
\newblock Synthetic dimension in photonics.
\newblock {\em Optica}, 5(11):1396--1405, November 2018.
\newblock Publisher: Optica Publishing Group.

\bibitem{fang_realizing_2012}
Kejie Fang, Zongfu Yu, and Shanhui Fan.
\newblock Realizing effective magnetic field for photons by controlling the phase of dynamic modulation.
\newblock {\em Nature Photonics}, 6(11):782--787, November 2012.
\newblock Publisher: Nature Publishing Group.

\bibitem{tholen_measurement_2022}
Mats~O. Tholén, Riccardo Borgani, Giuseppe~Ruggero Di~Carlo, Andreas Bengtsson, Christian Križan, Marina Kudra, Giovanna Tancredi, Jonas Bylander, Per Delsing, Simone Gasparinetti, and David~B. Haviland.
\newblock Measurement and control of a superconducting quantum processor with a fully integrated radio-frequency system on a chip.
\newblock {\em Review of Scientific Instruments}, 93(10):104711, October 2022.

\bibitem{yamamoto_principles_2016}
Yoshihisa Yamamoto and Kouichi Semba, editors.
\newblock {\em Principles and {Methods} of {Quantum} {Information} {Technologies}}, volume 911 of {\em Lecture {Notes} in {Physics}}.
\newblock Springer Japan, Tokyo, 2016.

\bibitem{rivera_hernandez_control_2024}
J.~C. Rivera~Hernández, Fabio Lingua, Shan~W. Jolin, and David~B. Haviland.
\newblock Control of multi-modal scattering in a microwave frequency comb.
\newblock {\em APL Quantum}, 1(3):036101, July 2024.

\bibitem{naaman_synthesis_2022}
Ofer Naaman and José Aumentado.
\newblock Synthesis of {Parametrically} {Coupled} {Networks}.
\newblock {\em PRX Quantum}, 3(2):020201, May 2022.
\newblock Publisher: American Physical Society.

\bibitem{ranzani_circulators_2019}
Leonardo Ranzani and Jose Aumentado.
\newblock Circulators at the {Quantum} {Limit}: {Recent} {Realizations} of {Quantum}-{Limited} {Superconducting} {Circulators} and {Related} {Approaches}.
\newblock {\em IEEE Microwave Magazine}, 20(4):112--122, April 2019.
\newblock Conference Name: IEEE Microwave Magazine.

\bibitem{bock_data_2025}
Christoph~L. Bock, Juan~Carlos Rivera~Hernández, Fabio Lingua, and David~B Haviland.
\newblock Data {Repository} for the {Article} "{Non}-reciprocal {Scattering} in a {Microwave} {Frequency} {Comb}", {Zenodo}, \url{https://doi.org/10.5281/zenodo.15459741}, May 2025.
\newblock Version Number: 1.0.0.

\end{thebibliography}

\end{document}